\documentclass{paper_arxiv}

\title{Transferability and explainability of deep learning emulators for regional climate model projections: Perspectives for future applications}

\authors{Jorge Baño-Medina\aff{a}\correspondingauthor{Jorge Baño-Medina, bmedina@ifca.unican.es}
Maialen Iturbide,\aff{a} 
Jesús Fernández,\aff{a} 
José Manuel Gutiérrez,\aff{a} 
}

\affiliation{\aff{a}{Instituto de F\'isica de Cantabria (IFCA), CSIC-Universidad de Cantabria, Santander, Spain}}

\abstract{Regional climate models (RCMs) are essential tools for simulating and studying regional climate variability and change. However, their high computational cost limits the production of comprehensive ensembles of regional climate projections covering multiple scenarios and driving Global Climate Models (GCMs) across regions. RCM emulators based on deep learning models have recently been introduced as a cost-effective and promising alternative that requires only short RCM simulations to train the models. Therefore, evaluating their transferability to different periods, scenarios, and GCMs becomes a pivotal and complex task in which the inherent biases of both GCMs and RCMs play a significant role.
Here we focus on this problem by considering the two different emulation approaches proposed in the literature (PP and MOS, following the terminology introduced in this paper). In addition to standard evaluation techniques, we expand the analysis with methods from the field of eXplainable Artificial Intelligence (XAI), to assess  the physical consistency of the empirical links learnt by the models.
We find that both approaches are able to emulate certain climatological properties of RCMs for different periods and scenarios (soft transferability), but the consistency of the emulation functions differ between approaches. Whereas PP learns robust and physically meaningful patterns, MOS results are GCM-dependent and lack  physical consistency in some cases. Both approaches face problems when transferring the emulation function to other GCMs, due to the existence of GCM-dependent biases (hard transferability). This limits their applicability to build ensembles of regional climate projections. We conclude by giving some prospects for future applications. This work has been submitted to Artificial Intelligence for the Earth Systems. Copyright in this Work may be transferred without further notice.}

\begin{document}

\maketitle

%
%
%
\statement
Regional Climate Model (RCM) emulators are a cost-effective emerging approach for generating comprehensive ensembles of regional climate projections. Promising results have been recently obtained using deep learning models. However, their potential to capture the regional climate dynamics and to emulate other periods, emission scenarios, or driving Global Climate Models (GCMs) remains an open issue that affects their practical use. This study explores the potential of current emulation approaches incorporating new eXplainable Artificial Intelligence (XAI) evaluation techniques to assess the reliability and transferability of the emulators. Our findings show that the different global and regional model biases involved in the different approaches play a key role for transferability. Based on the results obtained we provide some prospects on potential applications of these models in challenging problems.

%
%
%

%

\section{Introduction}\label{sec1}

Regional\footnote{This work has been submitted to Artificial Intelligence for the Earth Systems. Copyright in this Work may be transferred without further notice.} Climate Models (RCMs, \cite{giorgi_thirty_2019}) are sophisticated tools widely used to produce  high-resolution regional climate projections. They work by numerically solving a set of physical equations representing regional atmospheric processes and interactions with other components, such as land, over a limited continental region. RCMs are driven at their boundaries by the coarse output of a Global Climate Model (GCM, \cite{phillips_general_1956}), a process often referred to as dynamical downscaling. A variety of studies \citep{rummukainen_added_2016, soares_simple_2018, molina_added_2022, cardoso_there_2022} have assessed the added value of RCMs, pointing to a better representation of the local scale as compared to their driving GCMs. Therefore, RCM simulations constitute a valuable line of evidence in assessing the risks and adaptation strategies related to climate change at the regional scale \citep{IPCC_2022}. 

The COordinated Regional climate Downscaling EXperiment (CORDEX) coordinates the generation of regional climate projections worldwide, based on multi-model ensembles of RCM simulations spanning different sources of uncertainty (including those arising from the driving GCM or the emission scenario, among others \cite{jacob_regional_2020, DiezSierra_cordex:2022}). However, covering the large number of potential scenario-GCM-RCM combinations is an enormous computational challenge. As a result, the limited availability of CORDEX simulations in some regions hinders the comprehensive assessment uncertainty in regional climate modeling \citep{kendon_using_2010, fernandez_consistency_2019}. This has led the regional climate modelling community to look for alternatives to these costly simulations, especially as they approach the km-scale.

Empirical-Statistical Downscaling (ESD) has traditionally been a cost-effective alternative/complement to dynamical downscaling \cite{maraun_statistical_2018}. ESD techniques rely on observations to learn the relationship between large-scale meteorological fields (typically from reanalysis datasets) and local surface variables of interest, such as temperature and precipitation. While the development of advanced machine learning techniques, such as deep Convolutional Neural Networks (CNNs, \citep{lecun_convolutional_1995}), opens up the potential for downscaling over large areas \citep{bano_suitability_2021}, the lack of sufficient observed data available for training in many regions remains as one of the main drawbacks of ESD that limits its applicability, particularly as finer spatial resolutions are considered.

An alternative approach recently introduced to overcome this problem is statistical RCM emulators. These models require an existing (relatively short) RCM simulation (driven by a particular GCM) to train the emulator, which learns the mapping between upper-air large-scale fields and surface target variables from the RCM. 
Two different RCM emulation approaches have been recently introduced in the literature, referred to as ``perfect'' and ``imperfect'' \cite{erlandsen_hybrid_2020, boe_simple_2022, hobeichi_using_2023}. Both use surface high-resolution RCM variables (typically temperature and/or precipitation) as target, or predictand, but  differ in the predictors used. The ``perfect'' approach uses a set of informative upscaled large-scale variables from the same RCM and can be therefore considered a hybrid implementation of the Perfect Prognosis (PP) downscaling approach, using observations (in this case pseudo-observations within the RCM world) for both predictors and predictand \cite{maraun_statistical_2018}. This approach maximizes the day-to-day correspondence and physical consistency between the input-output pairs, as both come from the same model. Here, we use well-established terminology in ESD and refer to this approach as ``PP'' (from perfect prognosis). On the contrary, the so-called `imperfect'' approach uses the driving GCM fields as predictors, thus coping with the lack of perfect day-to-day correspondence and model biases in the learning process. Similarly to the previous case, using standard terminology we refer to this approach as ``MOS'', in an analogy with the MOS (Model Output Statistics) downscaling approach, which deals with model biases during learning. 

Several studies have analyzed both emulation approaches independently \citep{doury_regional_2022, hobeichi_using_2023} or jointly \citep{boe_simple_2022, van_deep_2022}, based on the comparison of particular evaluation metrics/indices between the emulated and target RCM fields. These studies show promising results to emulate an intermediate temporal period for the same GCM-RCM pair, particularly for the MOS approach, since the PP one inherit GCM-RCM biases that affect the emulated fields. 
However, evaluating emulators is a challenging task due to the complexity (black box nature) of the underlying deep learning models, so it is important to be able to analyse the inner functioning of these models (e.g., to analyze the predictor-predictand patterns learned) to contextualise the results. Not including this type of analysis may lead to an incomplete assessment of the emulation capabilities of both PP and MOS approaches. Some recent studies have explored the application of eXplainable Artificial Intelligence (XAI) in conventional statistical downscaling tasks to e.g., assess the physical consistency of the predictor patterns used by the models for inference \citep{gonzalez_using_2023, bano_understanding_2020, balmaceda_use_2023, rampal_high-resolution_2022}. This new evaluation dimension allows for a better understanding of the capabilities and limitations of CNNs and may allow to interpret how the different GCM and RCM biases may affect both emulation approaches.

One of the most promising aspects of emulators is that they could be applied to complete the scenario-GCM-RCM matrix from partial simulations, ideally from a single scenario-GCM pair.
Ideally, the emulator should capture the regional climate dynamics of the GCM and be transferable. This means that it should be able to emulate other periods, emission scenarios or even driving GCMs than those considered in the training phase. However, this remains an open issue that affects the practical use of RCM emulators for climate change applications. 

In this work, we assess the transferability of PP and MOS statistical RCM emulators based on state-of-the-art CNNs. To this aim we combine both standard  and new XAI-based evaluation techniques, which allows us to measure the trustworthiness of the emulators while deepening into the understanding of the transferability of each approach to other time period or emission scenario (soft transferability), or driving GCM (hard transferability). Based on the current state of knowledge, we conclude by giving some perspectives for future applications of these methods in problems where they can facilitate progress as an alternative (or complementary) to RCMs.

\section{Data and Methods}

This study focuses on the Alps region (Figure \ref{fig:scheme}) to test whether the emulators can reproduce certain fine-scale atmospheric processes well reproduced by RCMs, such as the influence of orography or coastal temperature gradients.

Following previous work \citep{doury_regional_2022, boe_simple_2022}, we use simulations from the ALADIN63 RCM over Europe at a spatial resolution of 0.11$^\circ$, provided by EURO-CORDEX \citep{jacob_regional_2020} for the historical experiment (1980-2005) and RCP8.5 scenario (2006-2100), which span a wide range of climatic conditions. These simulations are available at the Earth System Grid Federation (ESGF) for a total of four driving GCMs: NorESM1-M \citep{bentsen_norwegian_2013}, CNRM-CM5 \citep{voldoire_cnrm-cm51_2013}, MPI-ESM-LR \citep{muller_higher-resolution_2018} and HadGEM2-ES \citep{bellouin_hadgem2_2011}. Here, we consider the first three, to which we will hereafter refer by the names of their respective modeling institutions for brevity: NorESM, CNRM, and MPI.

\subsection{Predictors and predictand}

As predictand, we use daily near-surface air temperature of ALADIN63 at the original 0.11$^\circ$ horizontal resolution, driven by the corresponding GCM under the historical and the RCP8.5 scenarios.

To establish the relationships between low-resolution predictors and high-resolution predictand variables, we exploit the advantage of deep learning algorithms to automatically select relevant features from high-dimensional predictor spaces \cite{bano_understanding_2020}. Here, we use a set of daily mean atmospheric variables typically used as predictors for downscaling purposes \citep{brands_well_2013, gutierrez_reassessing_2013, gutierrez_intercomparison_2019, bano_configuration_2020, quesada_repeatable_2022}: geopotential height (500, 700 hPa), specific humidity, air temperature and both zonal and meridional wind velocities at 3 different pressure levels (500, 700 and 850 hPa).

Predictor datasets (GCMs and upscaled RCM) were re-gridded by means of a first order conservative remapping to a common spatial resolution of 1.5$^\circ$. This leads to a 4-dimensional (time, latitude, longitude, variable) predictor field of dimensions $N \times 10 \times 13 \times 14$, with $N$ being the number of days in the training sample. In some experiments, we considered bias adjusted GCM predictors obtained by a simple monthly adjustment of mean values (relative to the upscaled RCM).

\subsection{Experimental framework}
\label{sec:expframe}

\begin{figure}[h!]
    \centering
    \includegraphics[width=\linewidth]{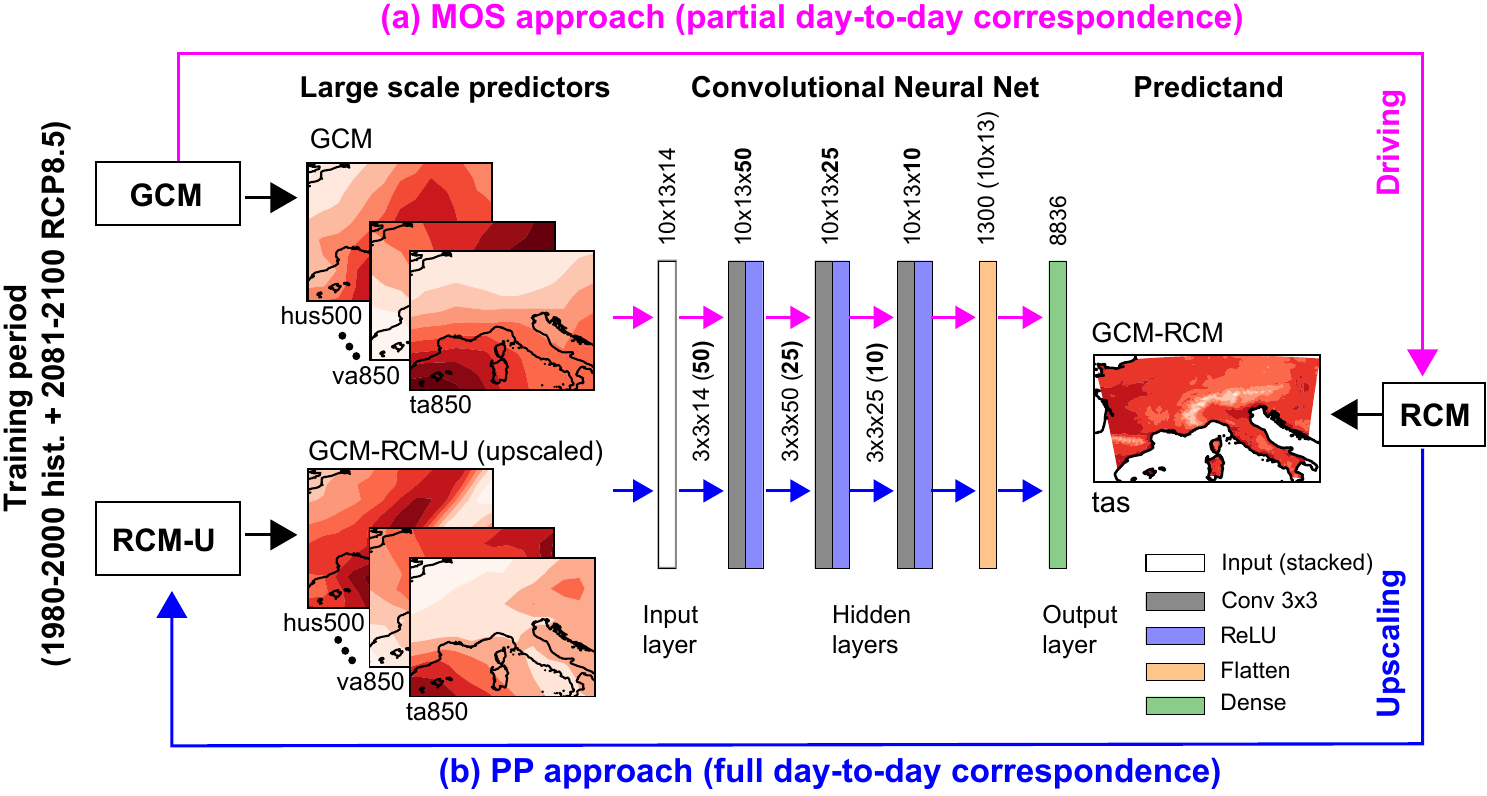}
    \caption{\label{fig:scheme}
    Schematic representation of the deep statistical emulator training workflows for the (a) MOS and (b) PP approaches. Arrows indicate the workflow. Details on the convolutional neural network model used are included in the figure: numbers on top of the convolutional layers represent their size, numbers between layers represent filter size and numbers in parenthesis indicate the number of filters used.
}
\end{figure}

We analyze and intercompare the two approaches introduced in the literature for statistical RCM emulation, referred to as ``perfect'' and ``imperfect'' \cite{boe_simple_2022}. Note that we refer to these approaches as ``PP'' and ``MOS'', respectively, based on the standard terminology used in statistical downscaling \cite{maraun_statistical_2018}. Figure~\ref{fig:scheme} provides a schematic illustration of these two approaches together with the Convolutional Neural Network (CNN) model used in this paper (see Section~\ref{s.deep}).

In the MOS emulator approach, predictors are taken directly from the driving GCM (Figure~\ref{fig:scheme}a, pink lines). The main shortcoming of this approach is the marginal correspondence between predictors and predictand fields (see Section~\ref{sec:correlations}), since the RCM is driven by the GCM at the boundaries of the domain, but develops its own dynamics within the domain. On the other hand, the PP emulator approach is trained with predictors obtained by upscaling the corresponding RCM fields \citep{doury_regional_2022}, so both predictors and predictand are physically consistent and have perfect day-to-day correspondence.
This is illustrated in Figure~\ref{fig:scheme}b (blue lines). 

In this study, we quantify and compare the strength of the day-to-day correspondence between the MOS and PP approaches. To do so, we computed the correlation of surface temperature at each RCM grid-box with the predictor fields for all predictor grid-boxes. Then, we selected the maximum value obtained in the spatial field, which typically corresponds to a nearby predictor grid-box. Correlations are computed separately for winter (DJF) and summer (JJA), partially eliminating the correlation arising solely from the annual cycle.

The experimental evaluation framework consists of three phases: training, evaluation and transferability. Building upon previous work by \cite{doury_regional_2022}, methods are trained using both the historical period (1996-2005, historical experiment) and a far-future period (2090-2099, RCP8.5 scenario) as depicted in Figure~\ref{fig:scheme}. As a result, we depart from the common practice of training in the historical period and testing in a future period. This allows us to separate the extrapolation capability related to the stationarity assumption from our analysis \citep{hernanz_critical_2022, doury_regional_2022}, which is actually more relevant for standard ESD methods, as they rely on observations to project into the future \citep{bano_downscaling_2022}.

\begin{figure}[!ht]
    \centering
    \includegraphics[width=\linewidth]{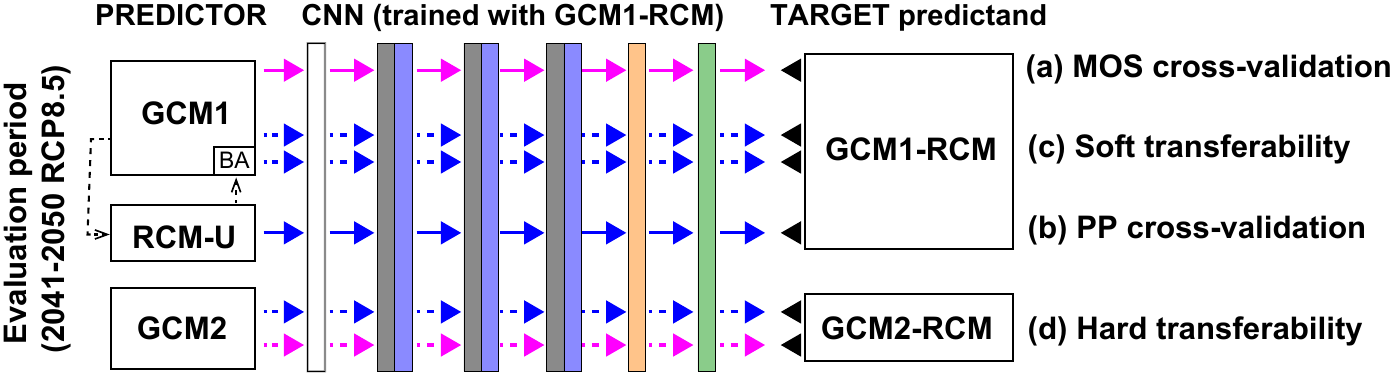}
    \caption{\label{fig:schemetest}
Experimental evaluation workflows to assess the performance of RCM emulators in (a-b) cross-validation and (c-d) transferability conditions for both PP and MOS approaches. Hard (inter-GCM) transferability (d) assesses the performance of RCM emulators when applied to predictors from different GCMs. Soft (intra-GCM) transferability (c) is special case for PP methods when applied to predictors from the same driving GCM used for the training phase; in this case, raw or bias adjusted (relative to the upscaled RCM) predictors from the GCM can be used. Arrow colors indicate the training approach (consistent with Figure~\ref{fig:scheme}: magenta for PP and blue for MOS), using solid for cross-validation and dashed for transferability.
 }
\end{figure}

In the evaluation phase, the methods trained with predictors from the GCM1 (MOS) or the upscaled GCM1-RCM (PP) are cross-validated by applying them to the same predictor sources, but from a different mid-future period (2041-2050; see Figure~\ref{fig:schemetest}a,b).

In the transferability phase (Figure~\ref{fig:schemetest}c,d), the performance of the trained methods is assessed by applying them to new predictor sources, arising either from the same driving GCM (here referred to as soft or intra-GCM transferability) or from different ones (hard or inter-GCM transferability).

Note that soft transferability is a special case for PP methods (which are trained using upscaled GCM-RCM predictors) to test their performance when applied directly to the GCM predictors of the same driving model used in the training phase. Large-scale discrepancies (biases) between GCM and upscaled RCM predictors may influence the emulated fields \citep{doury_regional_2022}. To overcome these discrepancies, there is the possibility to bias adjust (BA) the GCM fields using as reference the upscaled RCM variables \cite{boe_simple_2022}. Figure \ref{fig:schemetest}c illustrates the alternative use of raw or BA predictors with the two dashed lines. 

In hard transferability, the model trained with GCM1-RCM predictors is applied to a new GCM (GCM2) and, in principle, the emulator is expected to emulate the output of GCM2-RCM; note that this is challenging due to the different biases affecting the training and emulation phases (GCM1 and GCM2, respectively). 

\subsection{Deep learning models}\label{s.deep}

In this work we use Convolutional Neural Networks (CNNs) which are able to automatically infer complex spatial patterns from the input atmospheric spaces. In particular, we use the implementation proposed in \cite{bano_configuration_2020}, known as DeepESD. This model has been successfully used for downscaling purposes in the European continent for both precipitation and temperature fields \citep{bano_configuration_2020, bano_downscaling_2022}. Additionally, DeepESD has been analyzed with XAI techniques and proved able to learn plausible and coherent predictor-predictand links \citep{bano_understanding_2020}. 
Explainability is a key element of this study, used to understand the advantages and limitations of the two RCM emulator approaches.

DeepESD is a CNN composed of three convolutional layers (of 50, 25 and 10 filter maps, respectively) followed by a single dense layer; the particular configuration used is shown in Figure \ref{fig:scheme}. Each convolution is followed by a set of Rectified Linear Units (ReLU) to allow the statistical emulator to learn complex non-linear atmospheric predictor patterns. The  feature maps of the last hidden layer are flattened to build a dense connection with the output neurons, which correspond to the 8836 land gridpoints of ALADIN63 fields over the Alpine domain. The input layer is a stacked 4-D (variable, lon-lat grid, day) predictor field, which feeds the hidden structure of the CNN.

On the more technical side, we use an Adam optimizer \citep{kingma_adam_2014} to calibrate the model, with a batch size of 100, and a learning rate of 1E-4. We lean on a single NVIDIA Tesla V100 GPU with 32GB of memory to perform both calibration and inference.

We perform early-stopping with a patience of 30 epochs by randomly separating 10\% of the training data as our validation dataset to avoid overfitting. That is, when the loss in the validation dataset does not decrease in the next 30 epochs, then the network stops training. 

%
%
%
%
%
%
%
%
%
%
%
%
%
%

\subsection{Explainable artificial intelligence (XAI)}
\label{sec:xai}

Neural networks are seen as ``black-boxes" due to the complex operations occurring in their inner layers,  hindering interpretability and raising distrust in the results these models produce. To overcome this aspect, several XAI techniques have been recently developed to gain understanding about the underlying patterns inferred by deep neural networks. This is key to the task explored in this study, where the community aims to develop trustworthy and consistent RCM emulators to replace physics-based models.
Particularly, we lean on saliency maps, which are spatial representations of the relevance of the input features to the model predictions.

The relevance of a variable is predominantly measured by computing the gradients of the output space relative to the input space. These gradients are back-propagated through the hidden layers of the network and visually displayed in the form of saliency maps. Here, we follow previous work in climate-related applications \citep{kondylatos_wildfire_2022, gonzalez_using_2023}, and compute the saliency maps using the Integrated Gradients (IG) algorithm \citep{sundararajan_axiomatic_2017}. As a result, for each particular location in the predictand, the saliency maps have the same dimensions as the input features, which are 4-D arrays in our case.

In alignment with earlier studies \citep{toms_assessing_2021, mamalakis_explainable_2022} we post-process the ``raw'' gradients to enable a comparison across samples/days. To compute the saliency map of a particular day we follow five steps:

(1) we take the absolute value since we are interested in the relevance of the features regardless of their sign;
(2) we follow previous work \citep{toms_assessing_2021, mamalakis_explainable_2022, gonzalez_using_2023} and filter the lowest values to avoid gradient shattering by using a threshold of 1.5E-3;
(3) we compute the percentage for each saliency map individually by dividing the value of each feature by the total sum of all features;
(4) we filter again the gradients with a threshold of 1.5E-3 which now means 0.15\% relevance;
(5) we compute again the percentage of each saliency map.

Finally, we aggregate the saliency maps by averaging the values over the training period, resulting in a collection of maps, each representing a predictor variable. 

In this study, we illustrate the changing relevance patterns of the spatial predictors in four predictand grid-boxes located over France, the Southeastern Alps, Sardinia and Poland.

\section{Results}\label{sec:results}

\subsection{Predictor-predictand correlation}
\label{sec:correlations}

\begin{figure}[!ht]
    \centering
    \includegraphics[width=1\linewidth]{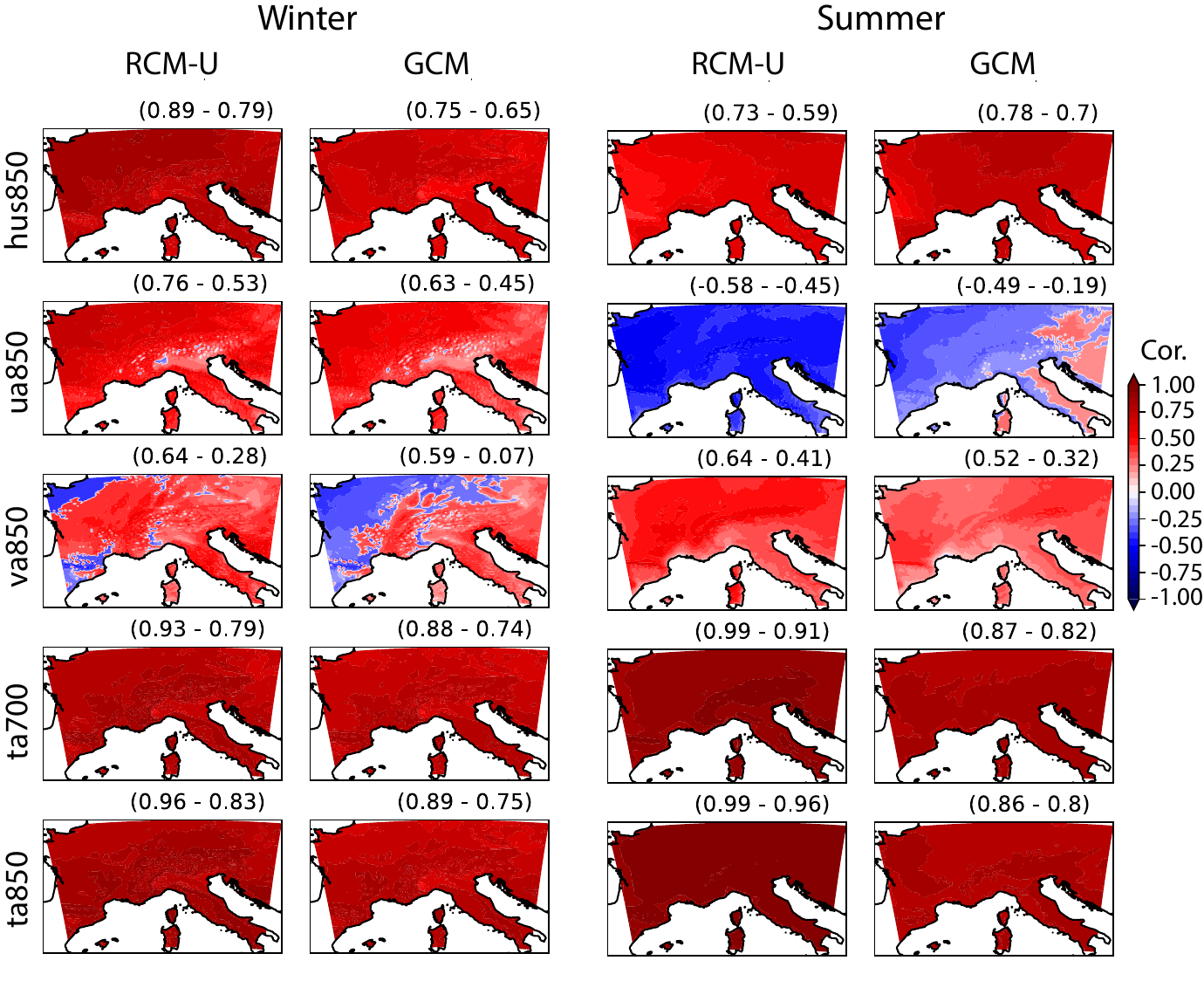}
    \caption{\label{fig:correlation}  Pearson temporal correlation of the daily series of  the NorESM-driven ALADIN63 surface temperature (predictand) and different predictors (in rows) from ALADIN63 (RCM-upscaled, RCM-U) and the NorESM GCM for winter (left) and summer (right). Values represent the maximum correlations obtained for each of the RCM predictand gridboxes (from the correlations with all  predictor gridbox values). The numbers on top of each panel indicate the maximum / mean spatial values.
    }
\end{figure}

The strength of the day-to-day correspondence between the predictors and the target predictand (RCM near-surface temperature) is depicted by the temporal correlation maps of Figure \ref{fig:correlation}, showing an illustrative example for a set of predictor variables and for the NorESM-driven ALADIN63 experiment (similar results are obtained for other driving models).


As expected, in general, upper-air, large-scale temperature fields show the highest correlation with near-surface temperature. Moisture shows also widespread high correlations, while winds show lower correlations and a seasonal pattern. This is consistent with e.g. westerlies (positive \textit{ua850}) advecting relatively warm air from the ocean in winter and cool air in summer.

Apart from the overall correspondence of the different predictors with the target variable, predictors from the GCM (second and fourth columns) show systematically lower correlations than those from the upscaled RCM (first and third columns). To aid in the comparison, the maximum correlation attained and the spatial mean are shown for each panel.
Moisture (\textit{hus850}) in summer is the only exception to the degradation of the predictor-predictand relationship in the GCM. The overall pattern with a higher correlation over northeastern Europe appears shifted to central Europe in the GCM predictors. 

These results reveal that, in general, the predictors from the PP approach are more informative for the deep emulator methods used in this work than those from the MOS approach. However, the latter may have some advantages since the CNN is learning directly the relationship between the driving GCM and the target RCM variable. In this way, the different biases can be accounted for directly, in analogy to the MOS approach for statistical downscaling \cite{maraun_statistical_2018}.  

\subsection{Cross-Validation}
\label{sec:crossval}
The different models are trained using historical (1996-2005) and far-future RCP8.5 (2090-2099) simulations, considering as predictors the upscaled RCM large-scale fields (PP approach) or the GCM fields (MOS approach) and using the corresponding RCM near-surface temperature as predictand. For the evaluation of the different approaches (Figure~\ref{fig:schemetest}a,b) we consider the mid-future RCP8.5 (2041-2050) period.

\begin{figure}[!ht]
  \centering
  \includegraphics[width=0.8\linewidth]{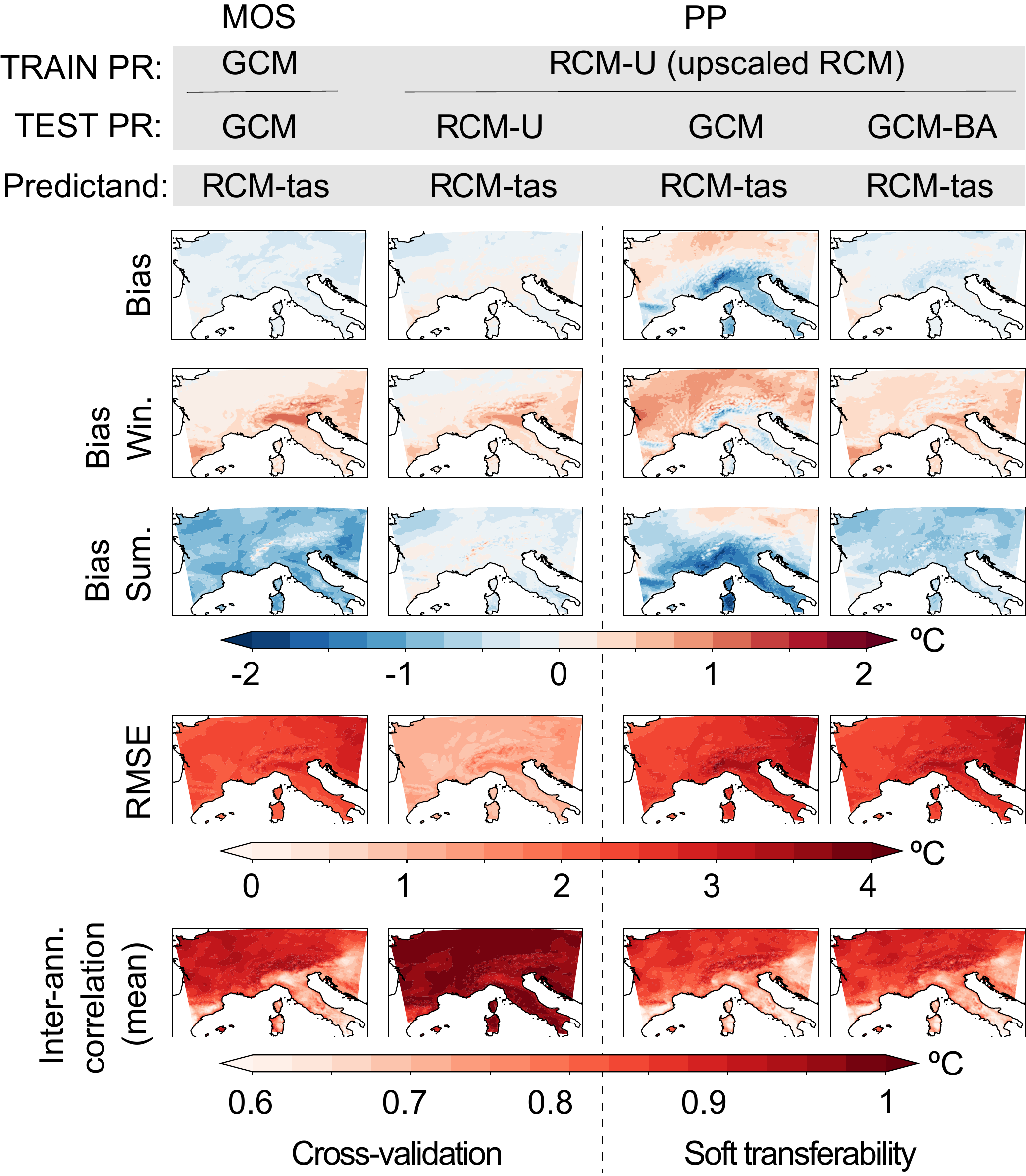}
  \caption{\label{fig:validation}
Evaluation metrics for emulated surface temperature over the test mid-future RCP8.5 (2041-2050) period for the ALADIN63 (RCM) simulations driven by the NorESM model (GCM). The first two columns show the cross-validated (same predictor source for train and test) results for the MOS and perfect approaches, respectively. The last two columns show the results corresponding to soft transferability for the perfect approach, using raw (third column) and bias adjusted (last column) GCM predictors; in this case, the predictors for the test period are taken from the driving GCM (instead of the upscaled RCM used for training). The labels on the top indicate the train and test predictors (PR)  and the predictands used in each case. The rows show, from top to bottom: annual, winter (DJF) and summer (JJA) biases, RMSE and interannual correlation. 
}
\end{figure}

Figure~\ref{fig:validation} shows the evaluation metrics (annual and seasonal biases, the RMSE and the interannual correlation, in rows) for emulated surface temperature over the test mid-future period for the ALADIN63 (RCM) simulations driven by the NorESM model (GCM). The first two columns show the cross-validation results for the MOS and PP approaches.

The PP approach attains lower RMSE values (around $1^\circ$C) than the MOS approach ($2-3^\circ$C). The former also reaches higher interannual correlation, with values above 0.95 north of the main mountain ranges and above 0.9 to the south. The MOS approach also shows a good correlation to the north (above 0.9) but the correlation drops to 0.7 to the south of the mountain barriers.

Annual biases are generally low but, for the MOS approach, they are the result of an average of larger seasonal biases of opposite signs in many locations. Note that the emulators are trained on an annual basis, including no specific predictor for the annual cycle. The largest winter warm biases occur in the Po Valley and are due to the differences in the mid-future test sample with respect to the training set, which are exacerbated particularly in the MOS approach.

\subsection{Soft Transferability of PP Methods}
\label{sec:soft}

The last two columns in Figure~\ref{fig:validation} show the results corresponding to soft transferability of the perfect approach to the driving GCM, using raw (third column) and bias adjusted (last column) predictors.

\begin{figure}[!ht]
    \centering
    \includegraphics[width=\linewidth]{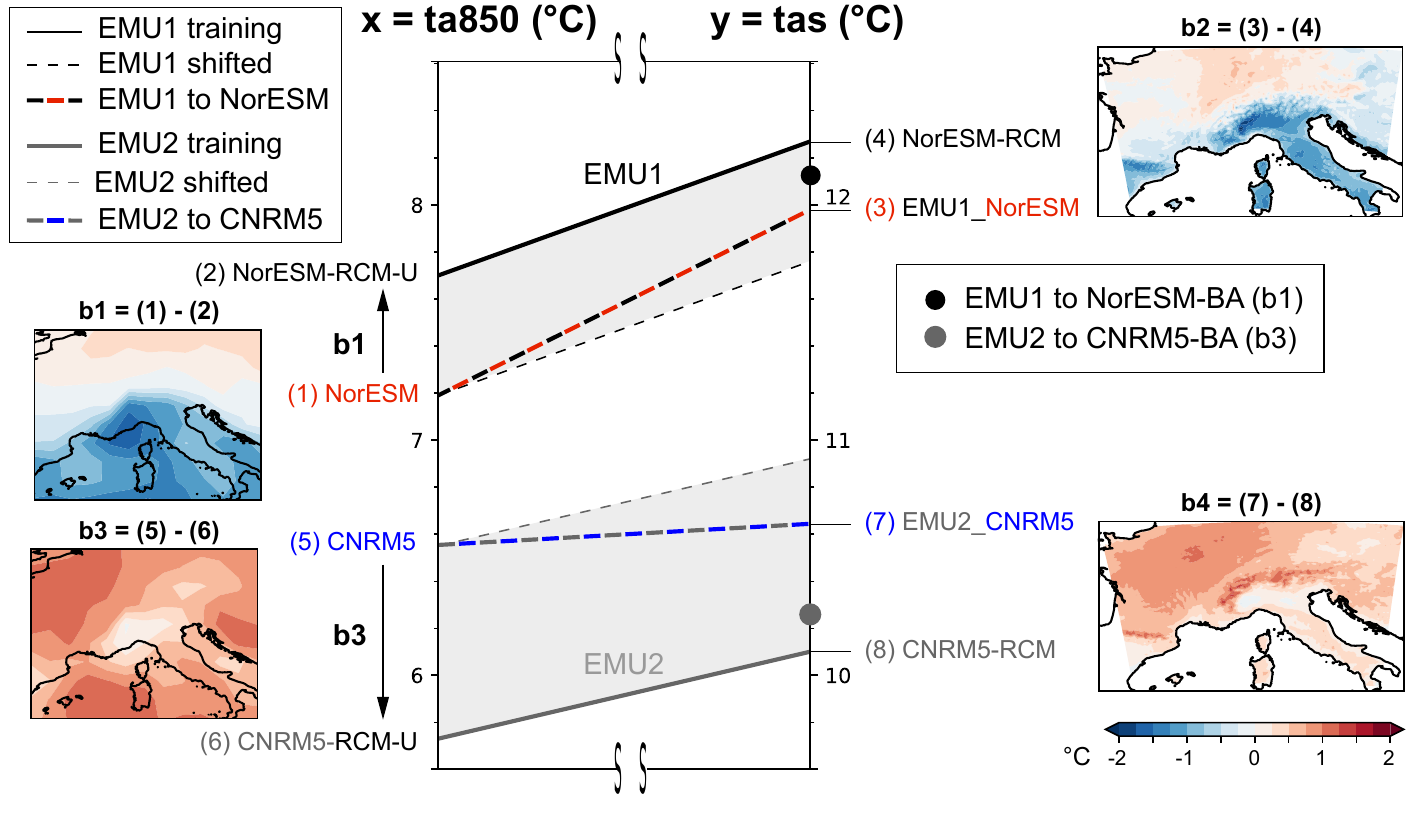}
    \caption{\label{fig:soft-transfer}
Soft transferability of the perfect approach for NorESM and CNRM GCM driving simulations. The left axis shows the spatial mean values of the large-scale temperature (the `ta850' predictor) for the upscaled RCM predictors used to train the emulators (RCM-U; 2 and 6) and their driving GCMs (1 and 5). The right axis shows the spatial means of the near-surface mean temperature (the `tas' predictand) resulting from the emulated models using GCM predictors (3 and 7) and the target RCM values (4 and 8). The thin dashed lines represent simple emulator extrapolated results for the GCM predictors (shifted EMU1 and EMU2). The different biases are indicated by b1-b4 notations and are accompanied by their corresponding spatial map representations. The results when bias-adjusted predictors are considered in the training phase are represented by solid black and grey dots on the right axis.
}
\end{figure}

The results show RMSE and interannual correlation patterns remarkably similar to the MOS approach, albeit with slightly higher RMSE and lower correlation. Notably, the annual and seasonal biases are significantly more pronounced when using the raw GCM fields as the test input (third column). As we show below (see Figure \ref{fig:soft-transfer}), this occurs due to the biases between the training upscaled RCM and their counterpart test GCM fields, as already identified in previous studies \cite{doury_regional_2022, boe_simple_2022}. Therefore, these biases could be alleviated by adjusting the GCM predictor biases (GCM-BA, relative to the upscaled RCM fields). Indeed, using GCM-BA as input (last column in Figure~\ref{fig:validation}), the emulator shows smaller biases, overall smaller than those of the MOS approach (in the first column). 

Qualitatively similar results are obtained for the emulators trained on the other GCM-driven simulations. We illustrate this through Figure~\ref{fig:soft-transfer} and the large-scale temperature (ta850), as is the most informative predictor in this study (see Sec. \ref{sec:correlations}). Note that for other variables, such as precipitation, the interpretation of the biases would be obscured by the combined effect of several relevant predictors on the target variable. The diagram of Figure~\ref{fig:soft-transfer} helps in better understanding the role that biases play, by representing linear transfer functions between two ordinate axes displaying the spatial averages of the predictor (ta850) and the predictand (tas) respectively. For instance, in the case of the NorESM-driven simulations, we observe the bias (b1) of the GCM (1) relative to the corresponding upscaled RCM predictor (2) that was used to train the EMU1 model, and how this bias is reflected in the bias of the predictand (b2) when comparing the resulting model output (3) with the target RCM values (4). Results for CNRM5-driven simulations are also shown in the same diagram (5-8 and b3-b4). Note that the two GCM predictors exhibit opposite biases when compared to the corresponding RCM upscaled predictors (b1 and b3). These biases are preserved to a large extent in the resulting predictions when comparing the emulated and actual RCM signals (b2 and b4, respectively). The maps on the sides display the corresponding spatial biases, reinforcing the idea that the large-scale biases in the predictors are inherited by the emulated surface temperature. Thus, these biases are reduced when adjusting the biases of the GCM predictors relative to the upscaled RCM fields, as indicated by the black and gray dots on the right `y' axis, corresponding to the results emulated from bias-adjusted GCM predictors, which are closer to the target RCM values.




These results illustrate the effect of model biases in emulators and show that the PP approach is suitable for emulating RCM outputs when using bias-adjusted predictors from the same driving GCM used for learning. This makes these models suitable for emulating RCM results in new periods and for new scenarios, at least when the new predictors fall into the range of variability used for learning.

\subsection{Hard transferability: Emulating from new driving GCMs}
\label{sec:transferability}

Hard transferability tests whether an emulator trained with a particular GCM-driven simulation can emulate the output that the RCM would have when driven by a different GCM. In this case, the model trained using data from GCM1-RCM is applied to a different GCM (GCM2) with the goal of reproducing the GCM2-RCM target (see Figure \ref{fig:schemetest}d). 
 
\begin{figure}[!ht]
    \centering
    \includegraphics[width=\linewidth]{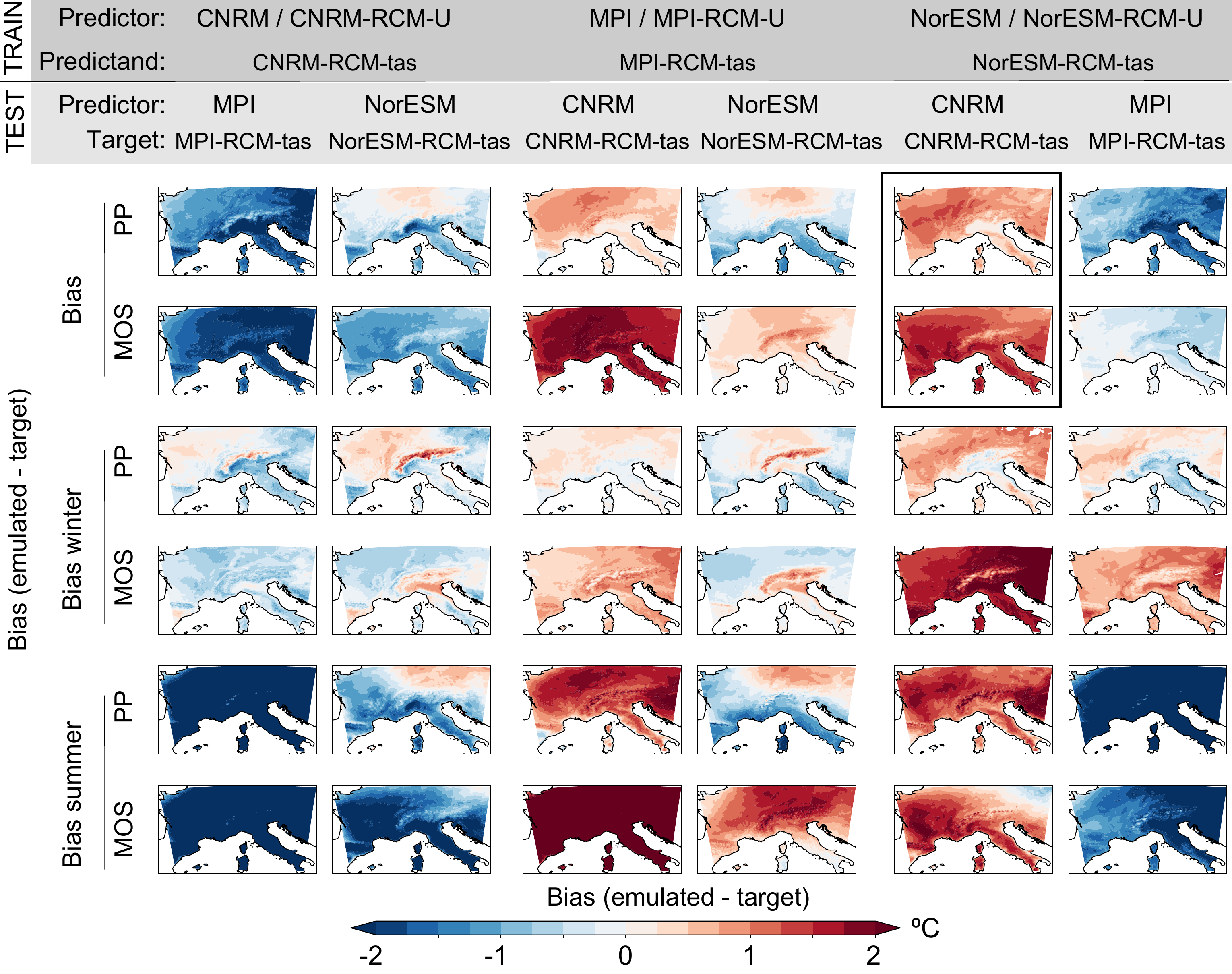}
    \caption{\label{fig:transfer_hard}
 Hard transferability of the perfect (EMU1) and imperfect (EMU3) approaches (EMU3) learnt with the NorESM driving simulation when applied to a different GCM (CNRM5).
    }
\end{figure}

Figure \ref{fig:transfer_hard} shows the biases resulting from hard transfer experiments (in columns) corresponding to the different combinations of different pairs of train and test GCMs from the set of available models (NorESM, CNRM and MPI). The first two rows show annual biases for the PP and MOS methods, respectively. The other rows show the corresponding seasonal results for winter and summer.
In most of the cases, the resulting biases exhibit a similar spatial pattern for the PP and MOS approaches, with smaller intensity for the former.
Also, the sign of the bias reflects the influence of the GCM on the emulated fields, with opposite biases when reversing the train-test GCMs (cf. columns 1-3, 2-5 and 4-6). This could reflect that the spatial structure of emulator biases is a consequence of the different biases of the GCMs used for training and prediction. 

In the case of soft transferability, the biases between the GCM and the upscaled RCM large-scale predictors were an avoidable source of error for RCM emulators. However, bias adjustment is not applicable for hard transferability, since adjusting GCM2 predictors relative to GCM1 would effectively yield a target output close to GCM1-RCM, instead of the desired target GCM2-RCM.

\begin{figure}[!ht]
    \centering
    \includegraphics[width=\linewidth]{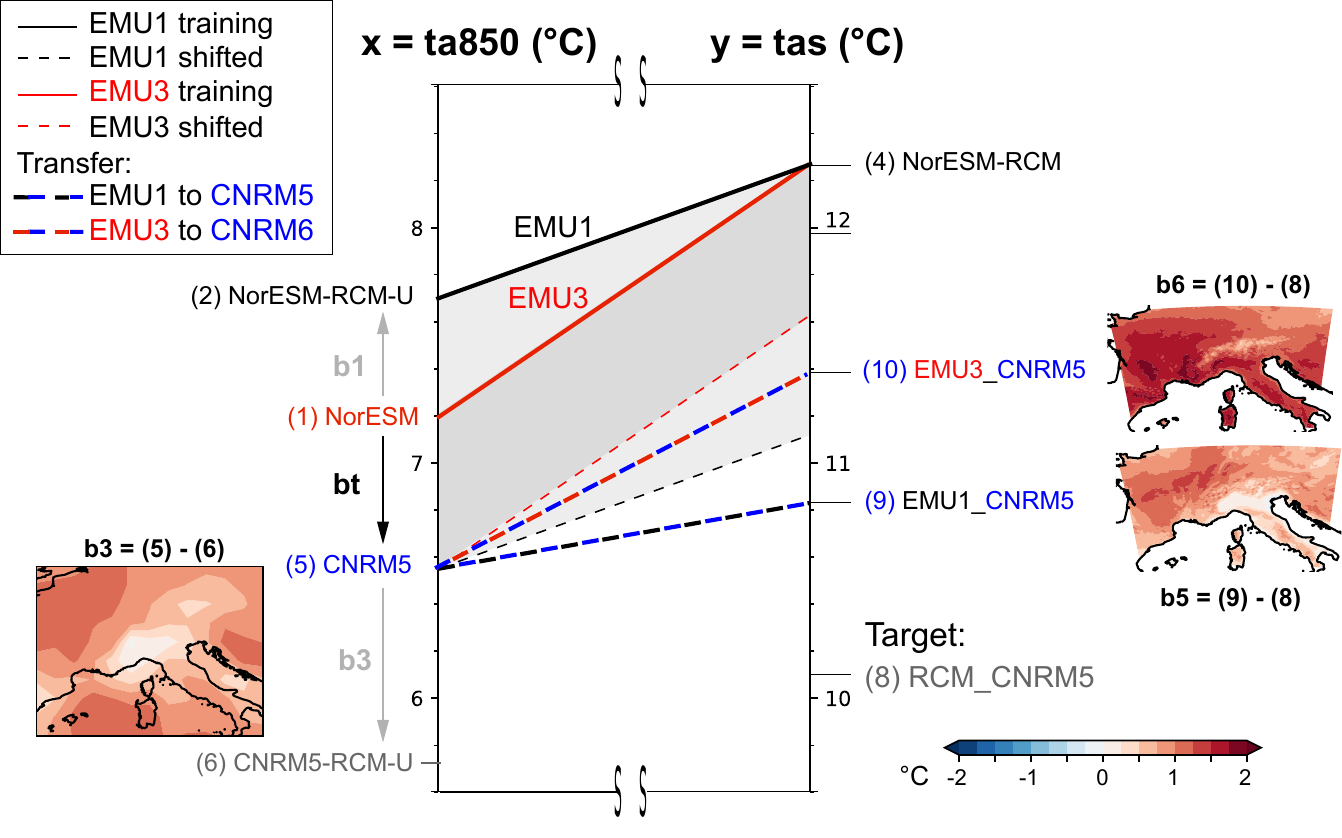}
    \caption{\label{fig:transfer1}
 Hard transferability of the PP (EMU1) and MOS (EMU3) emulator approaches learned with the NorESM driving simulation when applied to a different GCM (CNRM5).
    }
\end{figure}

The key here is that the biases between the upscaled RCM and the driving GCM can have opposite signs, as was shown in Figure~\ref{fig:soft-transfer} (b1 and b3). Thus, adjusting the biases across GCMs can be catastrophic for the emulator. Figure \ref{fig:transfer1} illustrates the effect of this hard transferability on the spatial average.
As in Figure~\ref{fig:soft-transfer}, emulators are depicted as linear transfer functions between two ordinate axes representing the spatial average ta850 (most informative predictor) and tas (predictand).
In this case, we also depict the MOS emulator (EMU3) trained with raw NorESM GCM predictors and NorESM-RCM tas.
On the spatial average, the emulators behave close to their linear representation, indicated in the plot by a simple shift of the emulator to the new test predictors (thin dashed lines), preserving the slope.
For instance, when applied to the CNRM GCM, the relatively strong slope of the MOS emulator (10) and the slighter one of the PP emulator (9), are preserved. But both positive slopes bring the emulator predictions far from the target (8).

Adjusting for the GCM bias would only worsen the results, since the emulator is already providing warm tas estimates using colder ta850 inputs from CNRM5. If these inputs are adjusted to the warmer NorESM fields, the emulator would provide even stronger warm tas biases.
For the PP emulator, the true adjustment to apply would be $b3$, that accounts for the difference between the upscaled RCM and the new GCM fields (CNRM5) provided as test input.
This adjustment would nudge the predicted fields in the right direction.
Note, however, that this adjustment cannot be done in practice, since the aim of this hard-transferability exercise is to avoid the RCM simulation nested into a second GCM. And, if this simulation is available, it is always more worth to train a new emulator (EMU2 in Figure~\ref{fig:transfer1}) on these new data than trying to adjust the inputs of an emulator (EMU1) trained on a different GCM. 

\subsection{Explainability}
\label{sec:explainability}
\begin{figure}[h!]
   \centering
   \includegraphics[width=\linewidth]{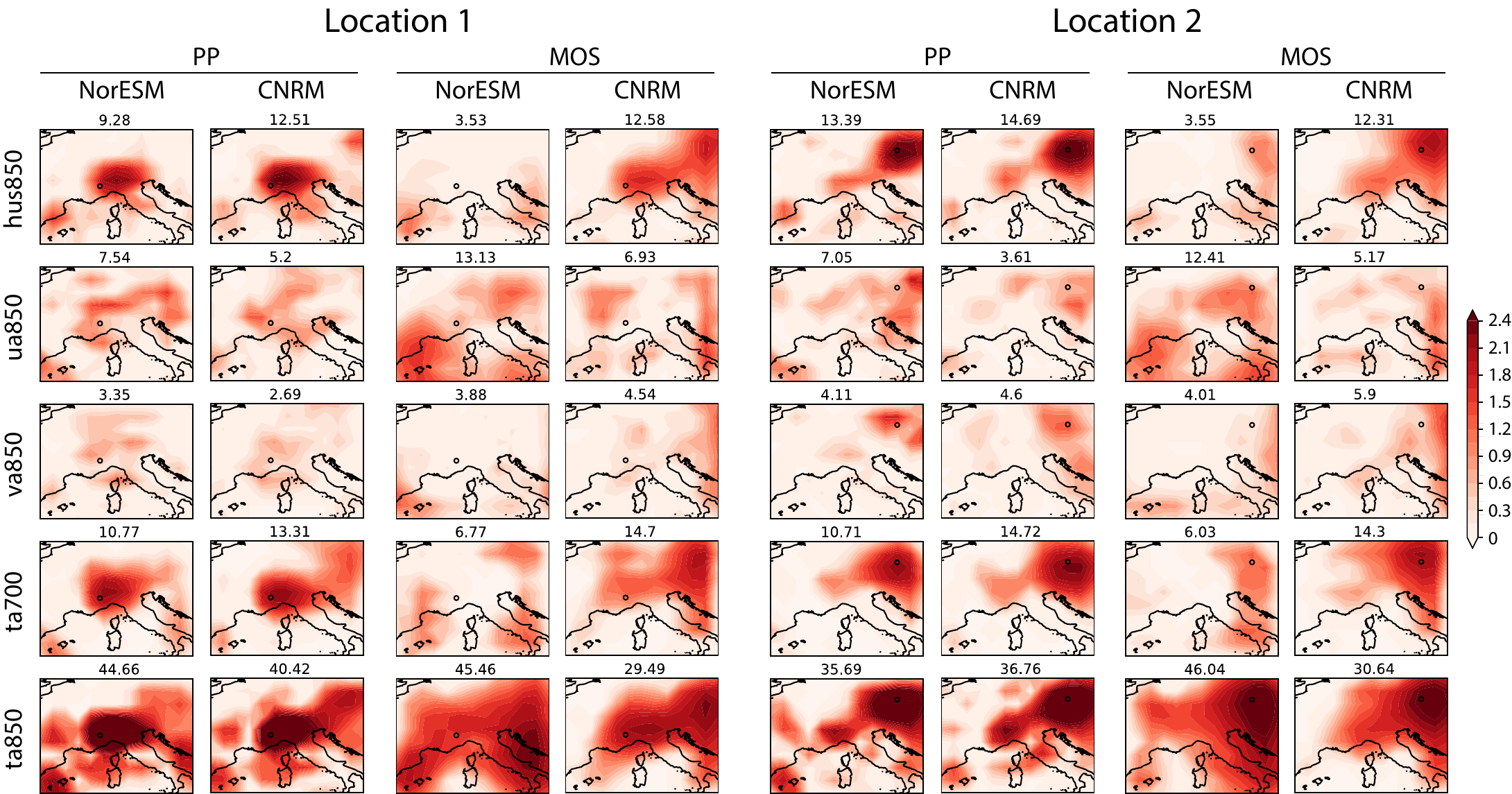}
   \caption{Aggregated (over the training period) saliency maps for winter temperature for two illustrative locations (loc1: close to the Alps and loc2: Poland; indicated by dots in each panel) for the PP and MOS models learnt for the NorESM and CNRM driving GCMs. Saliency values are displayed for  specific humidity (hus), zonal (ua) and meridional (va) wind velocities at 850 hPa, and air surface temperature (ta) at 700 and 850 hPa, in rows. The  numbers of each panel indicates the total contribution to the output (\%) of each predictor variable.}\label{fig:saliency1}
\end{figure}

Figure \ref{fig:saliency1} shows the aggregated (over the training period) saliency maps for two illustrative locations obtained for the PP and MOS deep models for two illustrative driving  GCMs (NorESM and CNRM). The figure displays the saliency maps for a selection of predictors: specific humidity (hus),  zonal (ua) and meridional (va) wind velocities at 850 hPa, and air surface temperature (ta) at 700 and 850 hPa, in rows. The  number of each panel indicates the total contribution to the output (\%) of each predictor variable. The saliency maps are displayed in pairs of columns, grouping the results for the two GCMs to facilitate the analysis of the patterns learnt when using different driving GCMs. Columns 1-2 (3-4) and 5-6 (7-8) show the PP (MOS) results for two illustrative locations (in the Alps and Poland), respectively.   

The figure shows remarkable differences in the relevance patters resulting from the two approaches. The PP results are very similar for the two GCMs, exhibiting a high local character, with patterns centered on the target location. This gives high confidence in the results, since the deep emulator is able to extract consistent patters of influence (with a clear physical interpretation) from different GCM-driven simulations of the same RCM. However, MOS patterns are in general more difficult to interpret, exhibiting non-localized or misplaced patterns that change from model to model.  In this example, the results for CNRM are in better correspondence with the patterns learnt with the PP approach, whereas the results for NorESM are in general more difficult to interpret, exhibiting non-localized or misplaced relevance patterns. This could be a consequence of the smaller day-to-day-correspondence between predictors and predictand for MOS, which could results in statistical artifacts with no physical consistency during the optimization process.  These results highlight the importance of introducing explainability in the evaluation of emulators, particularly for the MOS approach. The results for the MPI model (not shown) are more similar to the CNRM than the NorESM models. 

\begin{figure}[h!]
   \centering
   \includegraphics[width=\linewidth]{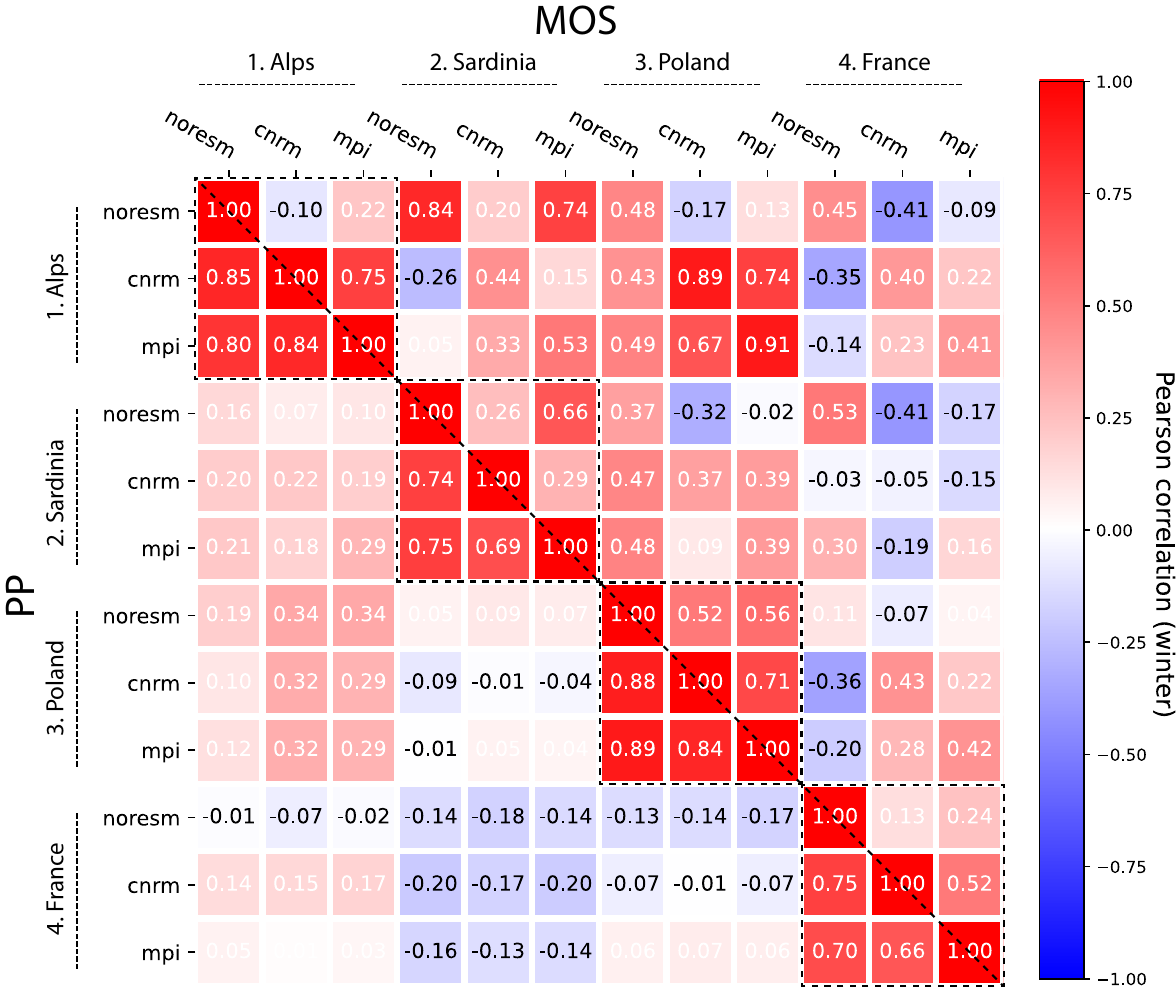}
   \caption{Correlation of saliency maps of humidity at 850 hPa for four illustrative locations for three RCM emulators build from different driving GCM (NorESM, CNRM and MPI). Values in the upper/lower triangle correspond to the results for the models learnt under the MOS/PP approaches (separated by the dashed diagonal). Dashed boxes along the diagonal show the inter-GCM results for each of the locations.}\label{fig:saliency2}
\end{figure}

These differences are further illustrated in Figure \ref{fig:saliency2} displaying the spatial correlations of the saliency maps for humidity at 850 hPa (the second most relevant predict, following upper-air temperatures) across different locations (four illustrative locations over the area of study) and three GCMs (NorESM, CNRM and MPI). The correlations for the different MOS/PP results are shown in the upper/lower triangles. This figure shows that the inter-GCM correlations for the PP approach are high for each of the locations, indicating similar spatial predictor patterns (from the upscaled GCM-driven RCM fields) being extracted by the different deep emulators (see the dashed boxes along the diagonal on the figure) for the same location. On the other hand, the PP results exhibit low inter-location correlations, indicating that the models learn specific spatial predictors for different locations. This is a desired behavior for the RCM emulators from a physical point of view.   

Contrarily, correlations from the MOS approach are in general medium and low, both across GCMs and across locations, indicating no apparent structure in the predictor fields relevant for the different models and locations. This is particularly relevant for the NorESM model, whereas  the MOS results for CNRM and MPI are closer to the PP results.  Therefore, there is no guarantee that RCM emulators trained under the MOS approach are able to extract meaningful physical information from the predictors. This requires a case by case assessment involving further research building on physical principles and processes. 

\section{Conclusions and Prospects}\label{sec4}

We applied RCM emulators using the two most common approaches (PP and MOS) to a set of existing GCM-RCM simulations from the CORDEX initiative.
We did so focusing on the soft and hard transferability of these emulators, evaluating them on the basis of physical consistency though an XAI technique (saliency maps).

Soft transferability would allow to confidently use these emulators on new test periods, new emission scenarios or new driving GCM members exploring natural climate variability.
We tested it with a new temporal period produced by the same training GCM-RCM pair.
Both PP, after adjusting the GCM predictors, and MOS successfully emulate unbiased temperature fields for summer and winter.
Nevertheless, XAI shows that PP learns from predictor patterns with a strong local dependence to the low-level temperature and humidity that resembles the actual climate dynamics.  
This agrees with studies dealing with conventional statistical downscaling (using observational data sets), suggesting similar local links for temperature \citep{gonzalez_using_2023}.
However, the predictability of MOS emulators leans on spatially extended, non-local patterns.
Moreover, these depend on the driving GCM used during training.
This difference between approaches is a result of the lower correlation between the input and output training fields in the MOS approach.
Under low day-to-day correspondence, the machine learning model is more likely to find predictability sources anywhere in the spatial domain. 

Hard transferability, that is, applying the emulator to predictor fields from unseen climates, was tested by using as input the fields from a new GCM, different from the training one.
In this case, the unique nature of the GCM-RCM combined biases prevent the emulator from being transferred.
Since the RCM response to the predictors can differ greatly, even in sign, between driving GCMs, the emulator cannot foresee this response in which it was not trained.
Further understanding of the nature of these biases would be needed, to identify transferability windows of opportunity for PP-based statistical emulators.
Recent work suggests that structural differences (e.g., aerosols representation and atmospheric physics) between the RCM and driving GCM are a large driver of the dissimilarity between their large-scale fields, being almost negligible when both simulations are driven with consistent external forcing among them \citep{taranu_mechanisms_2022}.
One such example, would be the CNRM-CM5-driven simulation, which indeed shows a bit more local predictor saliency patterns under MOS training than the other GCMs.

Overall, then, the PP approach learns the same robust predictor patterns for different driving GCMs thus increasing confidence.
These predictor patterns are locally connected to the predictand, resembling more closely the functioning of the RCM, that is ultimately the main goal.
On the other hand, MOS yields model-dependent patterns which, in some cases, are more difficult to interpret and lack from physical consistency.
This contrasts with previous studies, that recommended the use of MOS emulators against PP ones \citep{boe_simple_2022, van_deep_2022}, on the basis of conventional evaluation metrics.
We found that both approaches face problems in emulating the results for different driving GCMs, thus limiting their applicability to fill the GCM-RCM combination matrix of regional climate projections.

The limitations for the transferability of both RCM emulation approaches to be used on inputs from a new GCM (hard transferability) are difficult to overcome in general. Since they arise from the different response of the RCM to the driving GCM biases, so the way forward can be either to (1) train the emulator under non-biased driving fields, which could be used later also to de-bias new predictors or to (2) train the emulator with a wider variety of GCM biases.
The former could be accomplished by training with reanalysis data (so-called perfect boundary conditions for the RCM), but this would limit the training to the current climate, compromising the ability of the emulator to extrapolate under future climate conditions. The latter would imply mixing different driving GCMs in the training phase, in order to show the emulator the RCM response to different biases. This approach might enhance the capability of the emulator to cope with different biases, but it will likely affect its accuracy for any particular GCM input as compared to a specific GCM-RCM emulator as those shown in this work and elsewhere.

Emulators have also great potential for convection-permitting RCM (CP-RCM) simulations \citep{coppola_first_2020}.
These are very costly RCM simulations with grid spacings below 4km, where the parameterization representing convection can be deactivated, reducing uncertainties e.g. for precipitation.
These simulations are typically nested into the same RCM at coarser resolution (e.g. the 12 km resolution considered in this study).
In this context, the whole emulation process is performed in the framework of the same RCM, thus avoiding the problem of the biases arising from the driving model mismatch with the emulated one.
The emulator would learn the relationship between the RCM and CP-RCM and provide CP-RCM-emulated fields out of inputs from RCM simulations at coarse resolution.

Note that in this work we did not cover spatial transferability, i.e. training the model in one region (or on a representative selection of regions) and using it in different regions, which is also an active field of research \cite{Bjerre_2022, ludwig_assessing_2023}.

\clearpage
\acknowledgments
This work is part of IMPETUS4CHANGE, funded by the European Union's Horizon Europe research and innovation programme under grant agreement No 101081555. J.M.G and J.F acknowledge support from MCIN/AEI/10.13039/501100011033, which funded projects ATLAS (PID2019-111481RB-I00) and CORDyS (PID2020-116595RB-I00), respectively.

%
%
\datastatement
Both GCM and RCM data used in this study ---from the Coupled Model Intercomparison Project Phase 5 (CMIP5) and CORDEX, respectively,--- are openly available from the Earth System Grid Federation (ESGF, \url{https://esgf.llnl.gov}) portal. The code needed to implement the DeepESD model can be found in a Zenodo repository (\url{https://doi.org/10.5281/zenodo.6828303}, \cite{bano_medina_2022_zenodo}) associated to a previous publication \citep{bano_downscaling_2022}.

%






%



\bibliographystyle{paper_arxiv}
\bibliography{paper_arxiv}

\begin{thebibliography}{44}
\providecommand{\natexlab}[1]{#1}
\providecommand{\url}[1]{\texttt{#1}}
\renewcommand{\UrlFont}{\rmfamily}
\providecommand{\urlprefix}{URL }
\expandafter\ifx\csname urlstyle\endcsname\relax
  \providecommand{\doi}[1]{https://doi.org/\discretionary{}{}{}#1}\else
  \providecommand{\doi}{https://doi.org/\discretionary{}{}{}\begingroup
  \urlstyle{rm}\Url}\fi
\providecommand{\eprint}[2][]{\url{#2}}

\bibitem[{Balmaceda-Huarte et~al.(2023)Balmaceda-Huarte, Ba{\~n}o-Medina,
  Olmo,, and Bettolli}]{balmaceda_use_2023}
Balmaceda-Huarte, R., J.~Ba{\~n}o-Medina, M.~E. Olmo, and M.~L. Bettolli, 2023:
  On the use of convolutional neural networks for downscaling daily
  temperatures over southern south america in a climate change scenario.
  \textit{Climate Dynamics}, 1--15.

\bibitem[{Ba{\~n}o-Medina(2020)}]{bano_understanding_2020}
Ba{\~n}o-Medina, J., 2020: Understanding deep learning decisions in statistical
  downscaling models. \textit{Proceedings of the 10th International Conference
  on Climate Informatics}, 79--85.

\bibitem[{Ba{\~n}o-Medina et~al.(2022)Ba{\~n}o-Medina, Manzanas, Cimadevilla,
  Fern{\'a}ndez, Gonz{\'a}lez-Abad, Cofi{\~n}o,, and
  Guti{\'e}rrez}]{bano_downscaling_2022}
Ba{\~n}o-Medina, J., R.~Manzanas, E.~Cimadevilla, J.~Fern{\'a}ndez,
  J.~Gonz{\'a}lez-Abad, A.~S. Cofi{\~n}o, and J.~M. Guti{\'e}rrez, 2022:
  Downscaling multi-model climate projection ensembles with deep learning
  (deepesd): contribution to cordex eur-44. \textit{Geoscientific Model
  Development}, \textbf{15~(17)}, 6747--6758.

\bibitem[{Ba{\~n}o-Medina et~al.(2020)Ba{\~n}o-Medina, Manzanas,, and
  Guti{\'e}rrez}]{bano_configuration_2020}
Ba{\~n}o-Medina, J., R.~Manzanas, and J.~M. Guti{\'e}rrez, 2020: Configuration
  and intercomparison of deep learning neural models for statistical
  downscaling. \textit{Geoscientific Model Development}, \textbf{13~(4)},
  2109--2124.

\bibitem[{Ba{\~n}o-Medina et~al.(2021)Ba{\~n}o-Medina, Manzanas,, and
  Guti{\'e}rrez}]{bano_suitability_2021}
Ba{\~n}o-Medina, J., R.~Manzanas, and J.~M. Guti{\'e}rrez, 2021: On the
  suitability of deep convolutional neural networks for continental-wide
  downscaling of climate change projections. \textit{Climate Dynamics},
  \textbf{57~(11)}, 2941--2951.

\bibitem[{Baño-Medina et~al.(2022)Baño-Medina, Manzanas, Cimadevilla,
  Fernández, González-Abad, Cofiño,, and
  Gutiérrez}]{bano_medina_2022_zenodo}
Baño-Medina, J., R.~Manzanas, E.~Cimadevilla, J.~Fernández,
  J.~González-Abad, A.~S. Cofiño, and J.~M. Gutiérrez, 2022: {Repository
  supporting the results presented in the manuscript on Downscaling Multi-Model
  Climate Projection Ensembles with Deep Learning (DeepESD): Contribution to
  CORDEX EUR-44}. Zenodo.

\bibitem[{Bellouin et~al.(2011)}]{bellouin_hadgem2_2011}
Bellouin, N., and Coauthors, 2011: The hadgem2 family of met office unified
  model climate configurations. \textit{Geoscientific Model Development},
  \textbf{4~(3)}, 723--757.

\bibitem[{Bentsen et~al.(2013)}]{bentsen_norwegian_2013}
Bentsen, M., and Coauthors, 2013: The {Norwegian} {Earth} {System} {Model},
  {NorESM1}-{M} – {Part} 1: {Description} and basic evaluation of the
  physical climate. \textit{Geoscientific Model Development}, \textbf{6~(3)},
  687--720.

\bibitem[{Bjerre et~al.(2022)Bjerre, Fienen, Schneider, Koch,, and
  Højberg}]{Bjerre_2022}
Bjerre, E., M.~N. Fienen, R.~Schneider, J.~Koch, and A.~L. Højberg, 2022:
  Assessing spatial transferability of a random forest metamodel for predicting
  drainage fraction. \textit{Journal of Hydrology}, \textbf{612}, 128\,177,
  \doi{https://doi.org/10.1016/j.jhydrol.2022.128177},
  \urlprefix\url{https://www.sciencedirect.com/science/article/pii/S0022169422007508}.

\bibitem[{Boé et~al.(2022)Boé, Mass,, and Deman}]{boe_simple_2022}
Boé, J., A.~Mass, and J.~Deman, 2022: A simple hybrid statistical–dynamical
  downscaling method for emulating regional climate models over {Western}
  {Europe}. {Evaluation}, application, and role of added value? \textit{Climate
  Dynamics}, \doi{10.1007/s00382-022-06552-2},
  \urlprefix\url{https://doi.org/10.1007/s00382-022-06552-2}.

\bibitem[{Brands et~al.(2013)Brands, Herrera, Fern{\'a}ndez,, and
  Guti{\'e}rrez}]{brands_well_2013}
Brands, S., S.~Herrera, J.~Fern{\'a}ndez, and J.~M. Guti{\'e}rrez, 2013: How
  well do cmip5 earth system models simulate present climate conditions in
  europe and africa? \textit{Climate dynamics}, \textbf{41~(3)}, 803--817.

\bibitem[{Cardoso and Soares(2022)Cardoso, and Soares}]{cardoso_there_2022}
Cardoso, R.~M., and P.~M. Soares, 2022: Is there added value in the euro-cordex
  hindcast temperature simulations? assessing the added value using climate
  distributions in europe. \textit{International Journal of Climatology}.

\bibitem[{Coppola et~al.(2020)}]{coppola_first_2020}
Coppola, E., and Coauthors, 2020: A first-of-its-kind multi-model convection
  permitting ensemble for investigating convective phenomena over europe and
  the mediterranean. \textit{Climate Dynamics}, \textbf{55}, 3--34.

\bibitem[{Diez-Sierra et~al.(2022)}]{DiezSierra_cordex:2022}
Diez-Sierra, J., and Coauthors, 2022: The worldwide c3s cordex grand ensemble:
  A major contribution to assess regional climate change in the ipcc ar6 atlas.
  \textit{Bulletin of the American Meteorological Society}, \textbf{103~(12)},
  E2804 -- E2826, \doi{https://doi.org/10.1175/BAMS-D-22-0111.1},
  \urlprefix\url{https://journals.ametsoc.org/view/journals/bams/103/12/BAMS-D-22-0111.1.xml}.

\bibitem[{Doury et~al.(2022)Doury, Somot, Gadat, Ribes,, and
  Corre}]{doury_regional_2022}
Doury, A., S.~Somot, S.~Gadat, A.~Ribes, and L.~Corre, 2022: Regional climate
  model emulator based on deep learning: concept and first evaluation of a
  novel hybrid downscaling approach. \textit{Climate Dynamics}, 1--29.

\bibitem[{Erlandsen et~al.(2020)Erlandsen, Parding, Benestad, Mezghani,, and
  Pontoppidan}]{erlandsen_hybrid_2020}
Erlandsen, H.~B., K.~M. Parding, R.~Benestad, A.~Mezghani, and M.~Pontoppidan,
  2020: A hybrid downscaling approach for future temperature and precipitation
  change. \textit{Journal of Applied Meteorology and Climatology},
  \textbf{59~(11)}, 1793--1807.

\bibitem[{Fern{\'a}ndez et~al.(2019)}]{fernandez_consistency_2019}
Fern{\'a}ndez, J., and Coauthors, 2019: Consistency of climate change
  projections from multiple global and regional model intercomparison projects.
  \textit{Climate dynamics}, \textbf{52~(1)}, 1139--1156.

\bibitem[{Giorgi(2019)}]{giorgi_thirty_2019}
Giorgi, F., 2019: Thirty {Years} of {Regional} {Climate} {Modeling}: {Where}
  {Are} {We} and {Where} {Are} {We} {Going} next? \textit{Journal of
  Geophysical Research: Atmospheres}, \textbf{124~(11)}, 5696--5723,
  \doi{10.1029/2018JD030094},
  \urlprefix\url{https://onlinelibrary.wiley.com/doi/abs/10.1029/2018JD030094}.

\bibitem[{Gonz{\'a}lez-Abad et~al.(2023)Gonz{\'a}lez-Abad, Ba{\~n}o-Medina,,
  and Guti{\'e}rrez}]{gonzalez_using_2023}
Gonz{\'a}lez-Abad, J., J.~Ba{\~n}o-Medina, and J.~M. Guti{\'e}rrez, 2023: Using
  explainability to inform statistical downscaling based on deep learning
  beyond standard validation approaches. \textit{arXiv preprint
  arXiv:2302.01771}.

\bibitem[{Guti{\'e}rrez et~al.(2013)Guti{\'e}rrez, San-Mart{\'\i}n, Brands,
  Manzanas,, and Herrera}]{gutierrez_reassessing_2013}
Guti{\'e}rrez, J.~M., D.~San-Mart{\'\i}n, S.~Brands, R.~Manzanas, and
  S.~Herrera, 2013: Reassessing statistical downscaling techniques for their
  robust application under climate change conditions. \textit{Journal of
  Climate}, \textbf{26~(1)}, 171--188.

\bibitem[{Guti{\'e}rrez et~al.(2019)}]{gutierrez_intercomparison_2019}
Guti{\'e}rrez, J.~M., and Coauthors, 2019: An intercomparison of a large
  ensemble of statistical downscaling methods over europe: Results from the
  value perfect predictor cross-validation experiment. \textit{International
  journal of climatology}, \textbf{39~(9)}, 3750--3785.

\bibitem[{Hernanz et~al.(2022)Hernanz, Garc{\'\i}a-Valero, Dom{\'\i}nguez,, and
  Rodr{\'\i}guez-Camino}]{hernanz_critical_2022}
Hernanz, A., J.~A. Garc{\'\i}a-Valero, M.~Dom{\'\i}nguez, and
  E.~Rodr{\'\i}guez-Camino, 2022: A critical view on the suitability of machine
  learning techniques to downscale climate change projections: Illustration for
  temperature with a toy experiment. \textit{Atmospheric Science Letters},
  e1087.

\bibitem[{Hobeichi et~al.(2023)Hobeichi, Nishant, Shao, Abramowitz, Pitman,
  Sherwood, Bishop,, and Green}]{hobeichi_using_2023}
Hobeichi, S., N.~Nishant, Y.~Shao, G.~Abramowitz, A.~Pitman, S.~Sherwood,
  C.~Bishop, and S.~Green, 2023: Using machine learning to cut the cost of
  dynamical downscaling. \textit{Earth's Future}, \textbf{11~(3)},
  e2022EF003\,291.

\bibitem[{IPCC(2022)}]{IPCC_2022}
IPCC, 2022: \textit{Climate Change 2022: Impacts, Adaptation and
  Vulnerability}. Summary for Policymakers, Cambridge University Press,
  Cambridge, UK and New York, USA, 3-33 pp.

\bibitem[{Jacob et~al.(2020)}]{jacob_regional_2020}
Jacob, D., and Coauthors, 2020: Regional climate downscaling over europe:
  perspectives from the euro-cordex community. \textit{Regional environmental
  change}, \textbf{20~(2)}, 1--20.

\bibitem[{Kendon et~al.(2010)Kendon, Jones, Kjellstr{\"o}m,, and
  Murphy}]{kendon_using_2010}
Kendon, E.~J., R.~G. Jones, E.~Kjellstr{\"o}m, and J.~M. Murphy, 2010: Using
  and designing gcm--rcm ensemble regional climate projections. \textit{Journal
  of Climate}, \textbf{23~(24)}, 6485--6503.

\bibitem[{Kingma and Ba(2014)Kingma, and Ba}]{kingma_adam_2014}
Kingma, D.~P., and J.~Ba, 2014: Adam: A method for stochastic optimization.
  \textit{arXiv preprint arXiv:1412.6980}.

\bibitem[{Kondylatos et~al.(2022)Kondylatos, Prapas, Ronco, Papoutsis,
  Camps-Valls, Piles, Fern{\'a}ndez-Torres,, and
  Carvalhais}]{kondylatos_wildfire_2022}
Kondylatos, S., I.~Prapas, M.~Ronco, I.~Papoutsis, G.~Camps-Valls, M.~Piles,
  M.-{\'A}. Fern{\'a}ndez-Torres, and N.~Carvalhais, 2022: Wildfire danger
  prediction and understanding with deep learning. \textit{Geophysical Research
  Letters}, e2022GL099368.

\bibitem[{LeCun et~al.(1995)LeCun, Bengio et~al.}]{lecun_convolutional_1995}
LeCun, Y., Y.~Bengio, and Coauthors, 1995: Convolutional networks for images,
  speech, and time series. \textit{The handbook of brain theory and neural
  networks}, \textbf{3361~(10)}, 1995.

\bibitem[{Ludwig et~al.(2023)Ludwig, Moreno-Martinez, Hölzel, Pebesma,, and
  Meyer}]{ludwig_assessing_2023}
Ludwig, M., A.~Moreno-Martinez, N.~Hölzel, E.~Pebesma, and H.~Meyer, 2023:
  Assessing and improving the transferability of current global spatial
  prediction models. \textit{Global Ecology and Biogeography}, \textbf{32~(3)},
  356--368, \doi{10.1111/geb.13635},
  \urlprefix\url{https://onlinelibrary.wiley.com/doi/abs/10.1111/geb.13635}.

\bibitem[{Mamalakis et~al.(2022)Mamalakis, Ebert-Uphoff,, and
  Barnes}]{mamalakis_explainable_2022}
Mamalakis, A., I.~Ebert-Uphoff, and E.~A. Barnes, 2022: Explainable artificial
  intelligence in meteorology and climate science: Model fine-tuning,
  calibrating trust and learning new science. \textit{International Workshop on
  Extending Explainable AI Beyond Deep Models and Classifiers}, Springer,
  315--339.

\bibitem[{Maraun and Widmann(2018)Maraun, and
  Widmann}]{maraun_statistical_2018}
Maraun, D., and M.~Widmann, 2018: \textit{Statistical downscaling and bias
  correction for climate research}. Cambridge University Press, (2018).

\bibitem[{Molina et~al.(2022)Molina, Careto, Guti{\'e}rrez, S{\'a}nchez,, and
  Soares}]{molina_added_2022}
Molina, M.~O., J.~A.~M. Careto, C.~Guti{\'e}rrez, E.~S{\'a}nchez, and P.~M.~M.
  Soares, 2022: The added value of high-resolution euro-cordex simulations to
  describe daily wind speed over europe. \textit{International Journal of
  Climatology}.

\bibitem[{Müller et~al.(2018)}]{muller_higher-resolution_2018}
Müller, W., and Coauthors, 2018: A higher-resolution version of the {Max}
  {Planck} {Institute} {Earth} {System} {Model} ({MPI}-{ESM} 1.2-{HR}).
  \textit{Journal of Advances in Modeling Earth Systems}, \textbf{10}.

\bibitem[{Phillips(1956)}]{phillips_general_1956}
Phillips, N.~A., 1956: The general circulation of the atmosphere: A numerical
  experiment. \textit{Quarterly Journal of the Royal Meteorological Society},
  \textbf{82~(352)}, 123--164.

\bibitem[{Quesada-Chac{\'o}n et~al.(2022)Quesada-Chac{\'o}n, Barfus,, and
  Bernhofer}]{quesada_repeatable_2022}
Quesada-Chac{\'o}n, D., K.~Barfus, and C.~Bernhofer, 2022: Repeatable
  high-resolution statistical downscaling through deep learning.
  \textit{Geoscientific Model Development}, \textbf{15~(19)}, 7353--7370.

\bibitem[{Rampal et~al.(2022)Rampal, Gibson, Sood, Stuart, Fauchereau,
  Brandolino, Noll,, and Meyers}]{rampal_high-resolution_2022}
Rampal, N., P.~B. Gibson, A.~Sood, S.~Stuart, N.~C. Fauchereau, C.~Brandolino,
  B.~Noll, and T.~Meyers, 2022: High-resolution downscaling with interpretable
  deep learning: {Rainfall} extremes over {New} {Zealand}. \textit{Weather and
  Climate Extremes}, \textbf{38}, 100\,525,
  \doi{https://doi.org/10.1016/j.wace.2022.100525},
  \urlprefix\url{https://www.sciencedirect.com/science/article/pii/S2212094722001049}.

\bibitem[{Rummukainen(2016)}]{rummukainen_added_2016}
Rummukainen, M., 2016: Added value in regional climate modeling. \textit{Wiley
  Interdisciplinary Reviews: Climate Change}, \textbf{7~(1)}, 145--159.

\bibitem[{Soares and Cardoso(2018)Soares, and Cardoso}]{soares_simple_2018}
Soares, P.~M., and R.~M. Cardoso, 2018: A simple method to assess the added
  value using high-resolution climate distributions: application to the
  euro-cordex daily precipitation. \textit{International Journal of
  Climatology}, \textbf{38~(3)}, 1484--1498.

\bibitem[{Sundararajan et~al.(2017)Sundararajan, Taly,, and
  Yan}]{sundararajan_axiomatic_2017}
Sundararajan, M., A.~Taly, and Q.~Yan, 2017: Axiomatic attribution for deep
  networks. \textit{International conference on machine learning}, PMLR,
  3319--3328.

\bibitem[{Taranu et~al.(2022)Taranu, Somot, Alias, Bo{\'e},, and
  Delire}]{taranu_mechanisms_2022}
Taranu, I.~S., S.~Somot, A.~Alias, J.~Bo{\'e}, and C.~Delire, 2022: Mechanisms
  behind large-scale inconsistencies between regional and global climate
  model-based projections over europe. \textit{Climate Dynamics}, 1--26.

\bibitem[{Toms et~al.(2021)Toms, Barnes,, and Hurrell}]{toms_assessing_2021}
Toms, B.~A., E.~A. Barnes, and J.~W. Hurrell, 2021: Assessing decadal
  predictability in an earth-system model using explainable neural networks.
  \textit{Geophysical Research Letters}, \textbf{48~(12)}, e2021GL093\,842.

\bibitem[{van~der Meer et~al.(2022)van~der Meer, de~Roda~Husman,, and
  Lhermitte}]{van_deep_2022}
van~der Meer, M., S.~de~Roda~Husman, and S.~Lhermitte, 2022: Deep learning
  regional climate model emulators: a comparison of two downscaling training
  frameworks. \textit{Authorea Preprints}.

\bibitem[{Voldoire et~al.(2013)}]{voldoire_cnrm-cm51_2013}
Voldoire, A., and Coauthors, 2013: The {CNRM}-{CM5}.1 global climate model:
  description and basic evaluation. \textit{Climate Dynamics},
  \textbf{40~(9-10)}, 2091--2121.

\end{thebibliography}

\end{document}